\documentclass[12pt]{iopart}
\usepackage{epsf,epsfig,latexsym}
\usepackage{bm}
\usepackage{graphicx}
\usepackage{multirow}
\usepackage[rflt]{floatflt}
\usepackage{units}
\DeclareRobustCommand{\dgr}[1][]{
  \unit[#1]{\ifmmode{}^\circ\else${}^\circ$\fi}}

\def\ssNN#1{\sqrt{s_{NN}} \ifx|#1|\else=\unit[#1]{GeV}\fi}
\DeclareRobustCommand{\snn}[1]{\ifmmode\ssNN{#1}\else$\ssNN{#1}$\fi}
\DeclareRobustCommand{\valn}[3]{\ifmmode #1\,\pm\,_{#2\text{ (stat)}}^{#3\text{ (syst)}}\else$#1\,\pm\,_{#2\text{ (stat)}}^{#3\text{ (syst)}}$\fi}
\DeclareRobustCommand{\val}[3]{\ifmmode #1\,\pm\,_{#2}^{#3}\else$#1\,\pm\,_{#2}^{#3}$\fi}

\DeclareRobustCommand{\mpt}
{\ifmmode\left<p_T\right>\else$\left<p_T\right>$ \fi}
\DeclareRobustCommand{\pT}{\ifmmode p_T\else$p_T$\fi}
\DeclareRobustCommand{\mTm}{\ifmmode m_T-m\else$m_T-m$\fi}

\begin{document}

\title[Scanning the Phases of QCD]{Scanning the Phases of QCD with BRAHMS}

\author{Michael Murray for the BRAHMS Collaboration}
\address{University of Kansas, mjmurray@ku.edu, 785 864 3949}

\begin{abstract}
BRAHMS has the ability to study relativistic heavy ion collisions from the final freeze-out of hadrons 
all the way back to the initial wave-function of the gold nuclei. This is accomplished by studying hadrons with a very wide range of momenta and angles. 
 In doing so we can scan various phases of QCD, from a hadron gas, to a quark gluon plasma and perhaps to a color glass condensate. 
\end{abstract}



\section{Introduction}
The purpose of RHIC is to map the phase structure of QCD. 
So far the community has concentrated on AuAu, dAu and pp collisions at 
$\sqrt{s_{NN}}=200$ GeV 
in the hope of finding 
the quark gluon plasma. 
BRAHMS' special contribution 
has been 
to study the hadrons produced in these collisions over a broad range of 
$p_T$ and rapidity 
 \cite{BrNim}. 
A great deal of evidence supporting the creation of partonic matter in AuAu collisions was presented at this conference. However QCD is a rich theory that probably has many phases. It has been suggested that when viewed by a fast probe a heavy nucleus may 
resemble 
a sheet of highly correlated gluons called the Color Glass Condensate \cite{McLerranVenu}.
After a AuAu collision the fields generated by the color charges on the two sheets of gluons  
would break up into 
 partons which one would expect to approach chemical and kinetic equilibrium while rapidly expanding in both the longitudinal and transverse directions. 
Eventually the partons must hadronize and after further rescattering the hadrons freeze-out. 

This scenario is speculative and our evidence is both incomplete and somewhat indirect.   
Nevertheless in this paper 
we will attempt to map out this evolution by starting from the final state and working our way backwards.
The RHIC experiments have a beautiful complimentarity but we will report only on BRAHMS' data, with an emphasis on recent results. In particular we will discuss: 
\begin{itemize}
 \item Multiplicity distributions from dAu collisions \cite{BrdAuMult} which reflect  entropy production. 
 \item Particle spectra from AuAu \cite{DjamelQM04} which give information regarding;  
  \begin{itemize}
     \item kinetic freeze-out via blast wave fits to $p_T$ spectra,
     \item chemical freeze-out via fits to particle ratios, 
     \item initial pressure and longitudinal flow from pion dN/dy distributions. 
  \end{itemize}
  \item High $p_T$ suppression \cite{ZhangBao} 
which is sensitive to the early density of color charges.
   \item The ratio of  dAu and pp  spectra at high rapidities \cite{DebbeQM04} which   
give information on the Au wavefunction. 
\end{itemize}

\section{Global Observables}

\begin{figure}[h]
  \includegraphics[width=\columnwidth]{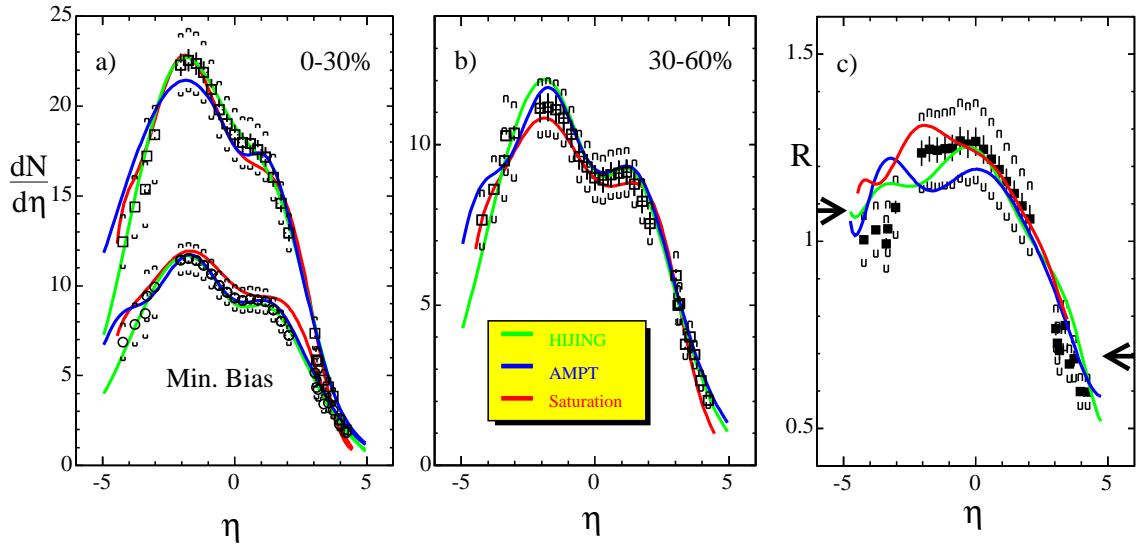}
  \caption{$dN/d\eta$ distributions from dAu collisions \cite{BrdAuMult} 
for a) minimum-bias and  0-30\% central events 
and b) 30-60\% central events. c) Scaled multiplicity/participant
ratio R. The left (right) arrows show
corresponding values for Au- (d-) participant scaling.  
}
  \label{dAuMult}
\end{figure}

Multiplicity distributions are sensitive to all stages of the collision and can be used to measure the total production of entropy. 
Figure~\ref{dAuMult} shows our $dN/d\eta$  results for minimum-bias and central dAu 
collisions \cite{BrdAuMult} (for AuAu see \cite{BrMult130, BrMult200}). Panel (c) shows the ratio of the
0-30\% and 30-60\% samples normalized by the number of participants. The ratios appropriate for Au- and d-participant only
scaling are 
indicated by the left and right arrows. 
Particle production away from 
mid-rapidity appears to follow the participant scaling of the respective
fragment. 
In the deuteron fragmentation region we see very similar yields to lower energy data. This phenomenon is known as ``limiting fragmentation" \cite{Benecke,BrdAuMult,BrMult200,PhobFrag}.  
The  HIJING and AMPT models are close to the data ~\cite{wang91,zhang01,lin01a}. Note that the saturation model results have been updated since the conference with a better 
centrality determination and an increase of the saturation scale from $Q^{2}_{s}  = 0.25$ to $0.34GeV^2$ \cite{kharzeev03}. These new calculations are close to the data.

\section{Particle Spectra}

\begin{figure}
\centering 
  \includegraphics[width=0.8\textwidth]{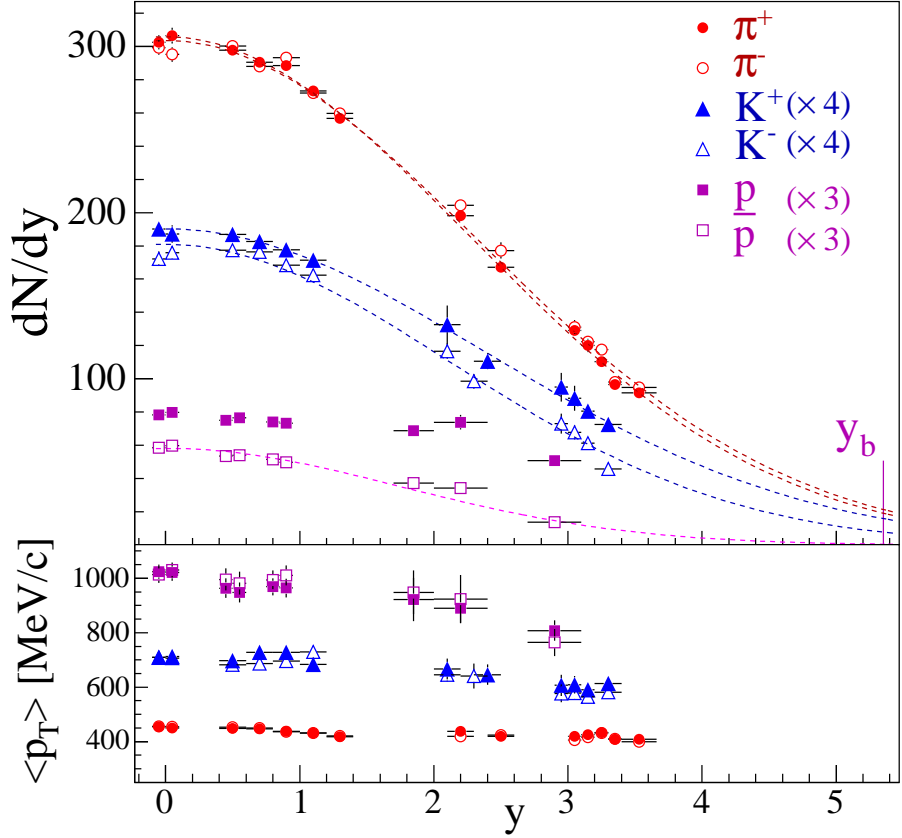}
  \caption{Rapidity densities (top) and  mean transverse momentum (bottom)
as a function of rapidity. Errors are statistical. 
The yields of all the produced particles are well described by Gaussian fits, shown by the dotted lines.
The beam rapidiity is 5.4}
  \label{dndy}
\end{figure}

The distribution of particles in rapidity and $p_T$ may give information on the transverse and longitudinal flow while the mix of different kinds of particles may tell us about the  ``quark chemistry" of the system. 
Our AuAu spectra are summarized in Fig.~\ref{dndy}, which shows the 
rapidity densities, dN/dy, and 
the mean transverse momenta, \mpt{}, for $\pi, K$, p and $\bar p$
 as a function of rapidity. Both quantities 
are estimated using fits to the spectra in narrow regions of rapidity \cite{DjamelQM04}.  
For $\pi, k$ and $\bar p$ 
the yields peak at y=0 and drop significantly at higher
rapidities.  The $\pi^+$ and $\pi^-$ yields are nearly equal
while an excess of $K^+$ over $K^-$ is observed that increases with rapidity.
The lower panel of Fig.~\ref{dndy} shows the rapidity dependence of \mpt{}.
There is no significant difference between particles and their antiparticles. In general, the rapidity
dependence of \mpt{} increases with mass suggesting that  transverse
flow drops with increasing rapidity. 
Using the net proton yield combined with baryon 
conservation and some assumptions on the neutron and hyperon yields allows us to estimate the total energy liberated by the stopping of the baryons.   We find $\Delta E= \int  EdN = 25 \pm 1$ TeV, or 75 GeV per participant \cite{DjamelQM04,peter}.

\subsection{Rapidity Dependence of Kinetic and Chemical Freeze-out}

\begin{figure}[ht]
  \begin{minipage}[b]{3.0in}
  \includegraphics[width=3.0in]{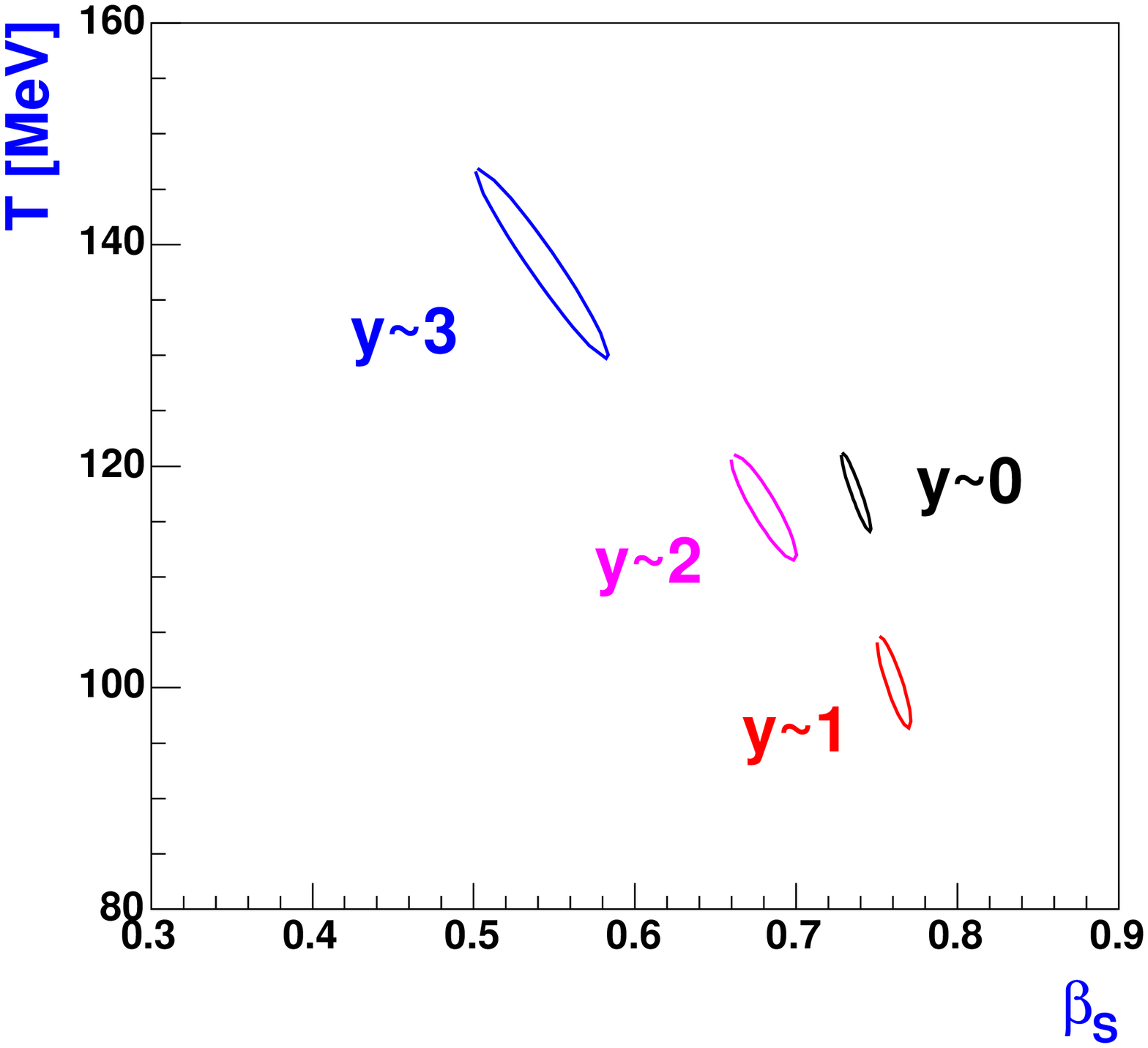}
   \end{minipage}
  \begin{minipage}[b]{3.0in}
  \includegraphics[width=3.0in]{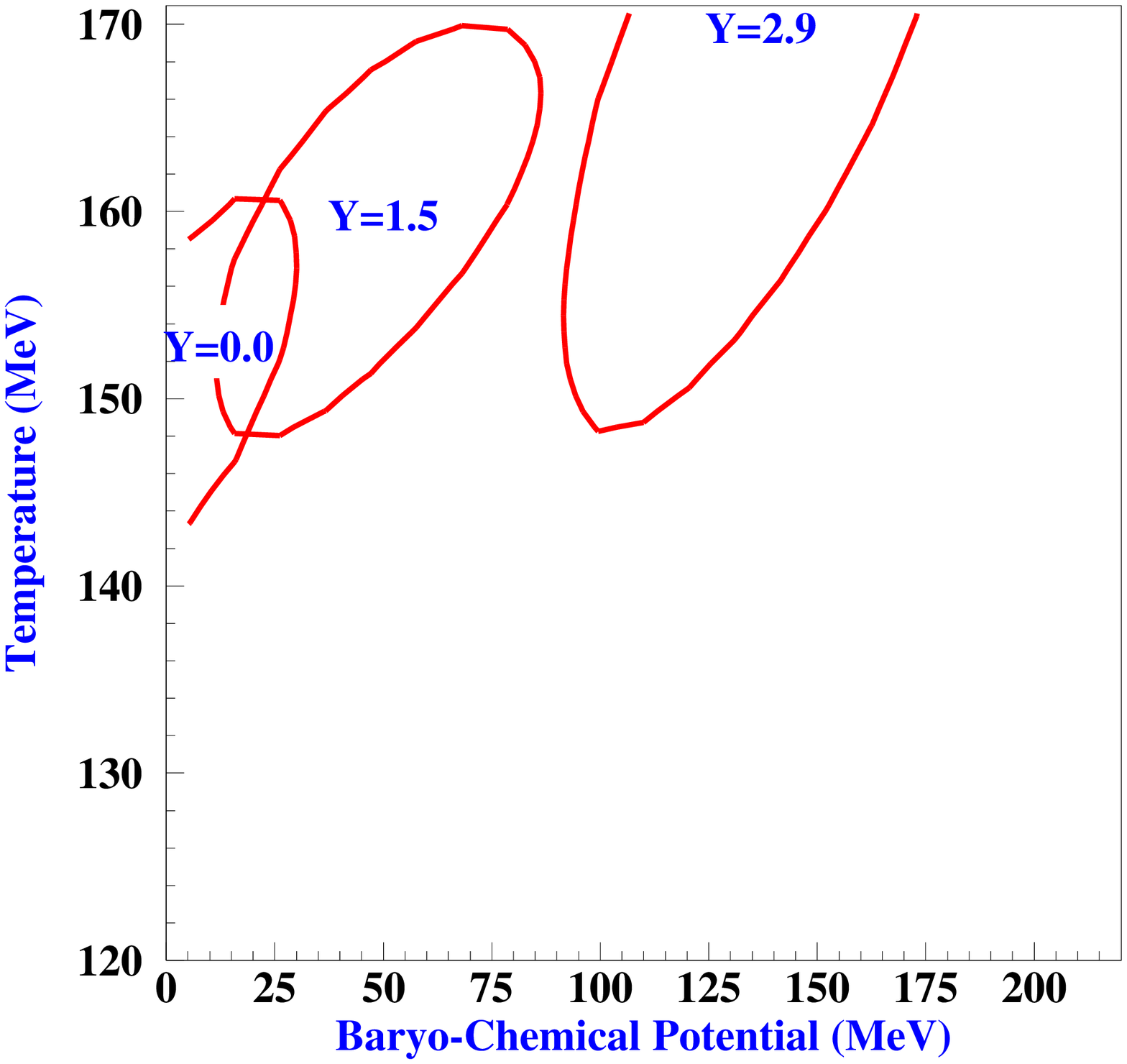}
   \end{minipage}
  \caption{Preliminary blast wave (left) and thermal (right) fits to our data at various rapidities. The curves are the one sigma contours. Note that as expected the thermal 
temperatures are always higher than the kinetic freeze-out temperatures.}
  \label{TvBeta}
\end{figure}

Is there one source or many in high energy heavy ion collisions?
We have investigated this question by fitting our spectra and particle yields 
at several different rapidities to blast wave and chemical models 
\cite{blast,Becattini}. 
At y=0 we see a very slow change of the freeze-out parameters with centrality so we shall consider only central data here. 
The left panel of 
Fig.~\ref{TvBeta} shows the regions of temperature T and transverse velocity of the surface $\beta_S$ that are consistent with our data sets at y=0,1,2 and 3. As the rapidity increases $\beta_S$ decreases while T increases. This may be because the equation of state of the matter is 
changing with rapidity. If the number of degrees of freedom decreases one would expect the temperature to increase.  
The right panel of Fig.~\ref{TvBeta} shows  
the results of a chemical analysis versus rapidity. 
As y increases both the baryo-chemical potential and (to a lesser extent) the chemical freeze-out temperature increase. Again this may suggest that the system has fewer degrees of freedom at higher rapidities.  
  
\subsection{Bjorken and/or Landau Hydrodynamics}
We now turn to the longitudinal flow which may be sensitive to the initial pressure in the system and possibly the equation of state. 
Bjorken proposed \cite{bjorken} that away from the fragmentation regions, $y\approx \pm 4$ at RHIC \cite{peter}, the system  produced by heavy ion collisions should be boost invariant, i.e. independent of rapidity.
This assumption is pervasive in the theoretical literature. Such an
expansion is the fastest possible one in the longitudinal direction. 
If the expansion is slower than the Bjorken limit then freeze-out occurs later and it may be easier to explain the fact that the pion  HBT radii  
are the same in the ``sidewards" and ``outwards" directions.

For $|y|<1$ all of our data are consistent with Bjorken's proposal.    
However looking globally at Fig.~\ref{dndy} there is a clear breakdown of boost invariance. 
This is most noticeable in the particle yields but it is also true that the \mpt 
of the kaons and antiprotons falls significantly with rapidity. 
Clearly a full understanding of the longitudinal dynamics would explain the  $\pi, k$ and $\bar p$ data. 
However because the pions dominate both the multiplicity and transverse energy, $E_T \equiv \sqrt{p_{T}^{2}+m^2}$, 
distributions 
focussing on the pions is a good 
start to 
a complete description of the longitudinal flow. Landau developed an analytic model of relativistic hydrodynamics undergoing an isentropic (constant entropy) 
expansion governed by an equation of state \cite{landau}. This 
idea 
was extended by Carruthers {\it et al} \cite{carut} to pion rapidity distibutions by assuming that 
pions are massless 
and that their $p_T$ and rapidity 
distributions approximately factorize. Under these conditions dN/dy is a gaussian with a width given by  
\begin{equation}
\sigma^2 = \ln{\left(\frac{\snn{}}{2m_N}\right)}
 \approx \ln \left({\gamma_{beam}} \right)
  \label{eq:width}
\end{equation}
where $m_N$ is the nucleon mass. 

\begin{figure}[htb] 
  \centering
  \includegraphics[width=\columnwidth]{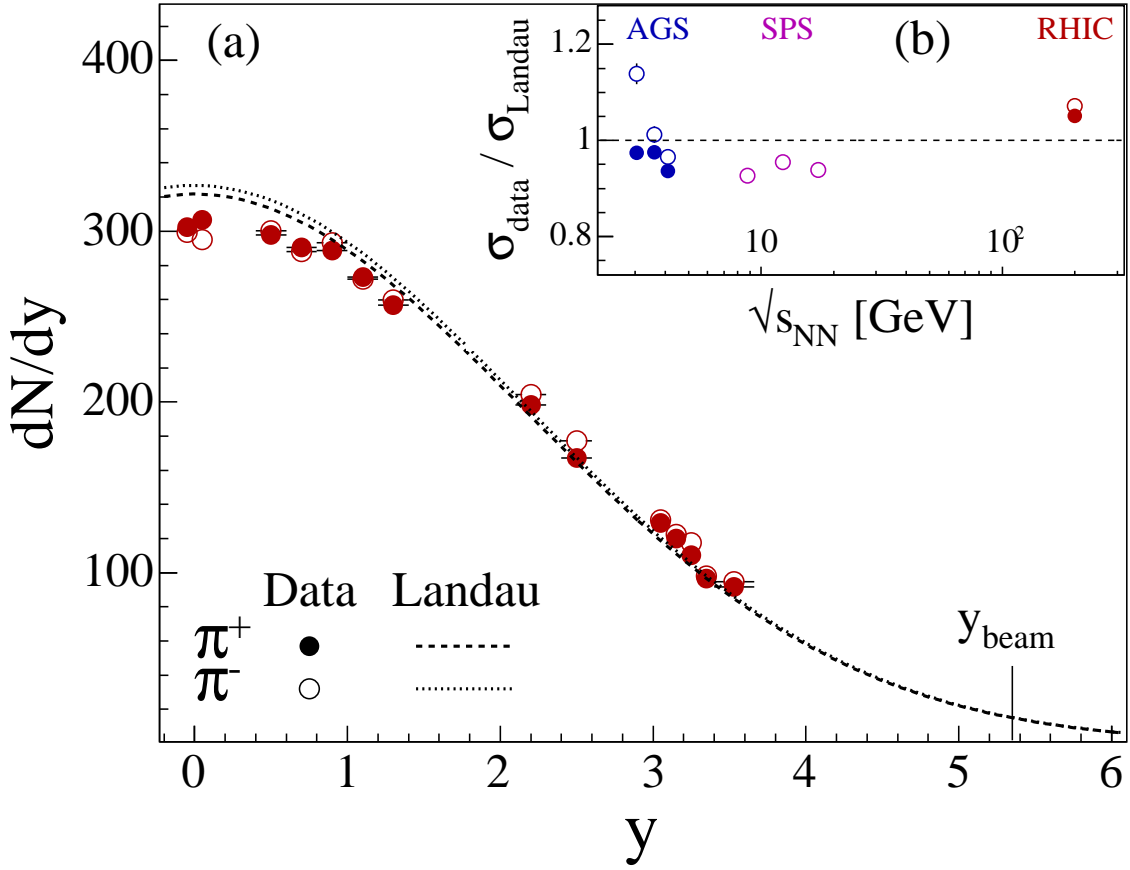}
  \caption{(a) Pion $dN/dy$ and Landau's Gaussian prediction at
    \snn{200} from Eq. \ref{eq:width}. The errors are statistical. (b)
    $\sigma_{N(\pi)}/\sigma_{Landau}$ as a function of $\snn{}$
    }
  \label{fig:landau}
\end{figure}

This model was able to give a reasonable description of the pion distributions from pp collisions at various energies. 
The  assumptions of the model are not entirely met for our data since $m_\pi = 0.3 \cdot$ \mpt at y=0 and \mpt drops by 10\% from y=0 to y=3. 
Another difference between our data and the Landau assumptions is that we do not observe full stopping. 
Nevertheless the agreement between this very simple model and our data is rather good.  
Figure~\ref{fig:landau}(a) shows $dN/dy(\pi)$ and Landau's prediction
for \snn{200} using Eq. \ref{eq:width} with the
condition that the integrals of these Gaussians must be equal to the
full--space yields estimated from the data.
A discrepancy of $\sim$ 5\% is observed ($\sigma_{Landau} =2.16$). 
Figure~\ref{fig:landau}(b)
shows a compilation on pion widths from AGS to RHIC, The
difference between theory and data is at most 10\%.  The logarithmic growth of the rapidity width with $\sqrt{s_{NN}}$ is in contrast to the linear increase of the multiplicity with $\sqrt{s_{NN}}$ \cite{PhobosRootS}. 
It is all the more striking considering that the degree of transparency
drastically changes from AGS to RHIC energies ~\cite{peter}. 

\section{High $p_T$ suppression and energy loss}
\begin{figure}[!tb]
  \includegraphics[width=1.0\columnwidth]{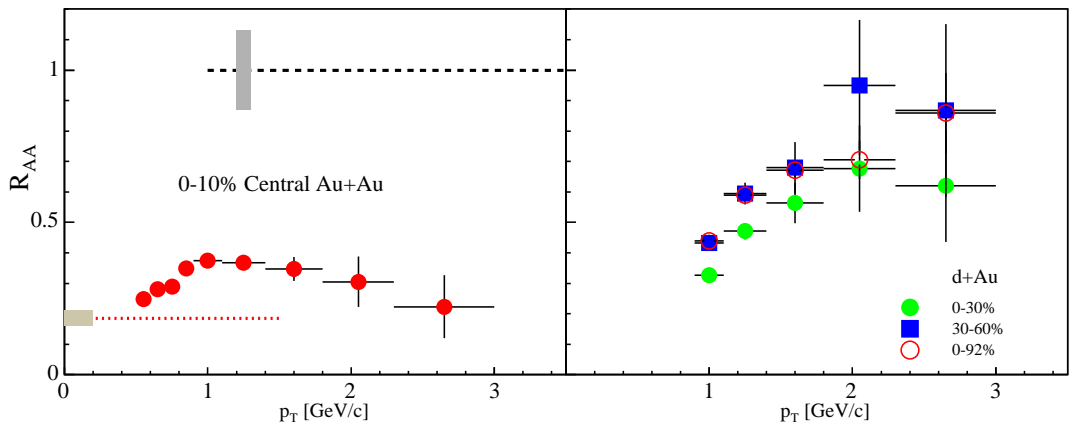}
  \caption{Preliminary $\pi^-$  nuclear modification factors at y=2.2 for AuAu 
and dAu collisions.  
The dashed and dotted lines represent the value expected 
if the yield scales by $\langle N_{coll} \rangle$ or $\langle N_{particpant} \rangle$  respectively. The
bands show our systematic errors on these quantities}
  \label{RAuAu22}
\end{figure}

The most exciting heavy ion news of 2003 was the discovery that
high $p_T$ suppression in AuAu collisions 
is a result of the hot and dense medium produced in these collisions 
rather than a depletion of hard partons in the Au nucleus itself  
\cite{HiPt4exp}. 
We quantify this effect by normalizing our spectra to pp
data 
using the nuclear modification factor; 
\begin{equation}
R_{AA}(p_T, y) \equiv \frac{1}{\langle N_{coll} \rangle}
        \frac{d^2N^{AuAu}/dp_T dy}{d^2N^{pp}_{inel}/dp_T dy}.
 \label{equation1}
\end{equation}
Here $\langle N_{coll} \rangle$ is the average number of nucleon-nucleon collisions in each event. BRAHMS has the unique ability to study this effect over a wide rapidity range \cite{ZhangBao}. 
Figure~\ref{RAuAu22} shows $R_{AA}$ for $\pi^-$   at y=2.2 for AuAu and dAu collisions. For central 
dAu collisions there is  already some suppression at y=2.2 although it is not as strong as in 
central AuAu collisons \cite{HiPt4exp,ZhangBao}. Note however that this measurement relies on 
an extrapolation for the pp reference based on PHENIX measurements, see \cite{ZhangBao}. 

\section{The Initial Gold Wavefunction} 
Finally it has been suggested that when viewed by a fast probe a heavy nucleus may form a new phase of QCD, the Color Glass Condensate ~\cite{McLerranVenu}. In such a state, two soft gluons will often  fuse together to produce one harder one. This effect results in a low $p_T$ suppression for dAu collisions compared to pp ones. 
BRAHMS can tune the speed of the probe by looking at fragments that emerge over a wide range of rapidities. 
Since we observe fragments close to the axis of the deuteron beam we are observing the initial parton density in the gold nucleus.
The partons momentum fraction, 
probed by a given
fragment is given by $x = e^{-y} p_T/\sqrt{s_{NN}}$. 
As y increases, x decreases and if the Color Glass scenario is correct we should 
see a shift of low $p_T$ particles to high $p_T$ with an overall reduction in yield. 
 
\begin{figure}[!htb] 
\centering
  \includegraphics[width=\columnwidth]{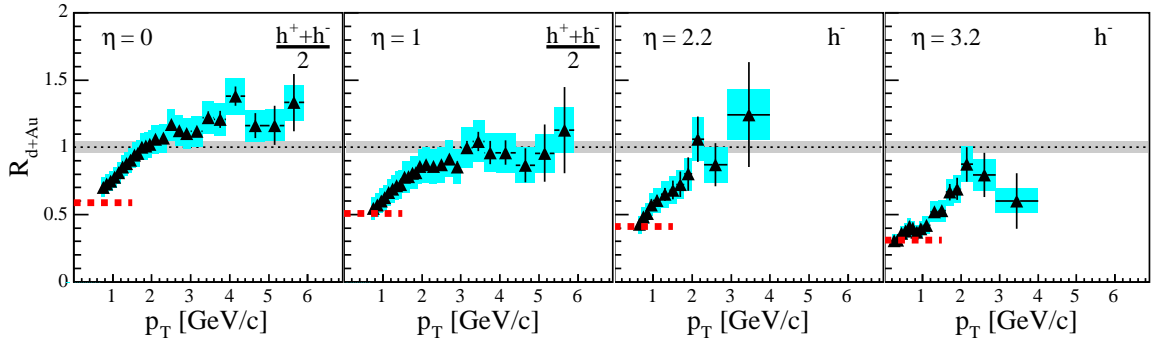}
\caption{\label{fig:ratio}Nuclear modification ratio for charged
  hadrons versus $p_T$ and  $\eta$.  Systematic
  errrors are shown with shaded boxes. The band around
  unity shows the error on 
$\langle N_{coll} \rangle$. 
  Dashed lines from  $p_T=0-1.5$ GeV/c show the 
ratio $\frac{1}{\langle
  N_{coll}\rangle}\frac{dN/d\eta(dAu)}{dN/d\eta(pp)}$.}
\end{figure}

Figure \ref{fig:ratio} shows $R_{dAu}$ for minimum--bias events as a function of $p_T$ and $\eta$. We use $\eta$ instead of 
y because we wish to combine all particles together and so improve our statistical accuracy. 
$R_{dAu}$ rises with $p_T$ and falls with $\eta$. 
The systematic errors in $R_{dAu}$ 
range from  $< 10-15$\%. 
 At midrapidity,
$R_{dAu}$ goes above 1. This so--called Cronin
enhancement ~\cite{Cronin} has been attributed to multiple scattering of the incoming 
partons  
during the collision. 
At $\eta=1$ the Cronin peak is not present and at
more forward rapidities ($\eta= 3.2$) the data show a suppression of
the hadron yields.  
At low $p_T$, $R_{dAu}$ is close to the ratio of charged-particle pseudorapidity 
densities in dAu and pp collisions 
~\cite{BrdAuMult,Alner:1986xu}.
Saturation effects should increase with the thickness of nuclear material 
traversed by the incoming probe and indeed  
we see a greater suppression for more central collisions, see Fig.~\ref{RAuAu22} and refs. \cite{BrRdAvY,DebbeQM04}. 

\section{Summary and Conclusions} 

For dAu collisions we see a significant asymmetry in 
$dN/d\eta$ with a peak at $\eta=-2$ (i.e. on the Au side of the collision) and a slight shoulder at $\eta=+2$. This suggests significant rescattering within the dAu system since these peaks are far away from the Au and d beam rapidities. In the fragmentation regions the multiplicity scales with the number of (the Au or deuteron) particpants. 
Using our spectra of identified particles we have found that the rapidity distributions of
all the produced charged particles in AuAu collisions are Gaussian. There is no large rapidity plateau but the data are consistent with boost invariance for $|y|<1$. The width of our 
pion distribution (and a large range of lower energy data) is consistent with Landau's picture of isentropic fluid dynamics. 

Blast wave analyses of 
our data show a decrease in the surface velocity $\beta_S$ and an increase in the kinetic freeeze-out temperature with increasing rapidity. Similarly chemical analysis of our particle yields hint that both the baryo-chemical potential and the chemical freeze-out temperature increase with rapidity. 
One could interpret this in terms of the system becoming less partonic (with consequently fewer degrees of freedom) at higher rapidities. 
The thermal analysis allows us to make a rough estimate of the total energy in the produced particles. This yields 
$25 \pm 5$ TeV compared to $25 \pm 1$ TeV computed from integrating the energy in the net protons \cite{peter}. 
 
We see evidence for jet quenching in AuAu collisions at both y=0 and y=2.2. The simplest  
explanation for these data is that fast partons lose energy traversing a deconfined system of quarks and gluons \cite{HiPt4exp}.  
In the forward region we see a low $p_T$ suppression of the yield from dAu collisions which increases with $\eta$ and centrality. 
This may be a result of a saturation in the yield of low momentum gluons in the  gold nucleus and hints that the Color Glass Condensate may represent the
high energy limit of QCD.

\section*{References}
\bibliographystyle{unsrt}

\end{document}